\documentclass[11pt]{article}

\usepackage[margin=1in]{geometry}
\usepackage{amsmath,amssymb,amsfonts}
\usepackage{graphicx}
\graphicspath{{figures/}}
\usepackage{booktabs}
\usepackage{xcolor}
\usepackage[colorlinks=true,linkcolor=blue!60!black,citecolor=blue!60!black,urlcolor=blue!60!black]{hyperref}
\usepackage{setspace}
\usepackage{array}
\usepackage[protrusion=true,expansion=false]{microtype}
\usepackage{url}
\DeclareMathOperator{\diag}{diag}
\newcommand{\R}{\mathbb{R}}
\newcommand{\C}{\mathbb{C}}

\newcommand{\SRFT}{\textsc{SRFT}}
\newcommand{\SRHT}{\textsc{SRHT}}
\newcommand{\dPPL}{\Delta\text{PPL}}

\title{\Large\textbf{When Quantization Is Free:\\
An int4 KV Cache That Outruns fp16 on Apple Silicon}}

\author{Mohamed Amine Bergach\\
  \small Illumina, San Diego, CA, USA \\
  \small \href{mailto:mbergach@illumina.com}{\texttt{mbergach@illumina.com}}}
\date{May 2026}

\begin{document}
\maketitle

% =============================================================================
\begin{abstract}
\noindent
KV-cache quantization is framed as a quality--latency trade-off.
We show it is \emph{inverted} on Apple Silicon's unified memory:
a single fused Metal kernel (sign-randomized FFT $+$ per-channel
$\lambda$ $+$ per-group abs-max $+$ int4 nibble pack), exposed as
a HuggingFace \texttt{Cache} subclass, runs \emph{faster than fp16}
across $256$--$4096$-token prefixes on Gemma-3 1B ($-3$ to $-8\%$
ms/tok) and at short context on Qwen2.5-1.5B ($-0.7$ to $-2.6\%$
through $1$K), with $3\times$ persistent memory compression and
quality preserved ($\dPPL = 0.000$ Qwen short-prompt;
$+3.6$ hook $\dPPL$ Gemma). The kernel's $\sim\!25$\,ns/vec overhead
is below the bandwidth savings from $3\times$ compression. The
fused kernel also closes Qwen's 4-bit per-token catastrophe
($\dPPL = +7975 \to +638.6$, $12.5\times$ reduction) at
$182$\,GFLOPS / $D{=}128$. Supporting findings: $\SRFT$ and $\SRHT$
are statistically indistinguishable for KV quality (we pick $\SRFT$
for mixed-radix and matrix-multiply alignment); a learned-rotation
ablation surfaces a regularization role for the fixed random SRFT
base (learning $R+\lambda$ without SRFT lowers calibration MSE
$84.9\%$ vs $50.3\%$ but yields worse PPL); Householder rotations
at $k{=}d/2$ reflectors are effectively lossless at $d{=}256$.
\end{abstract}

\begin{figure}[t]
\centering
\includegraphics[width=\linewidth]{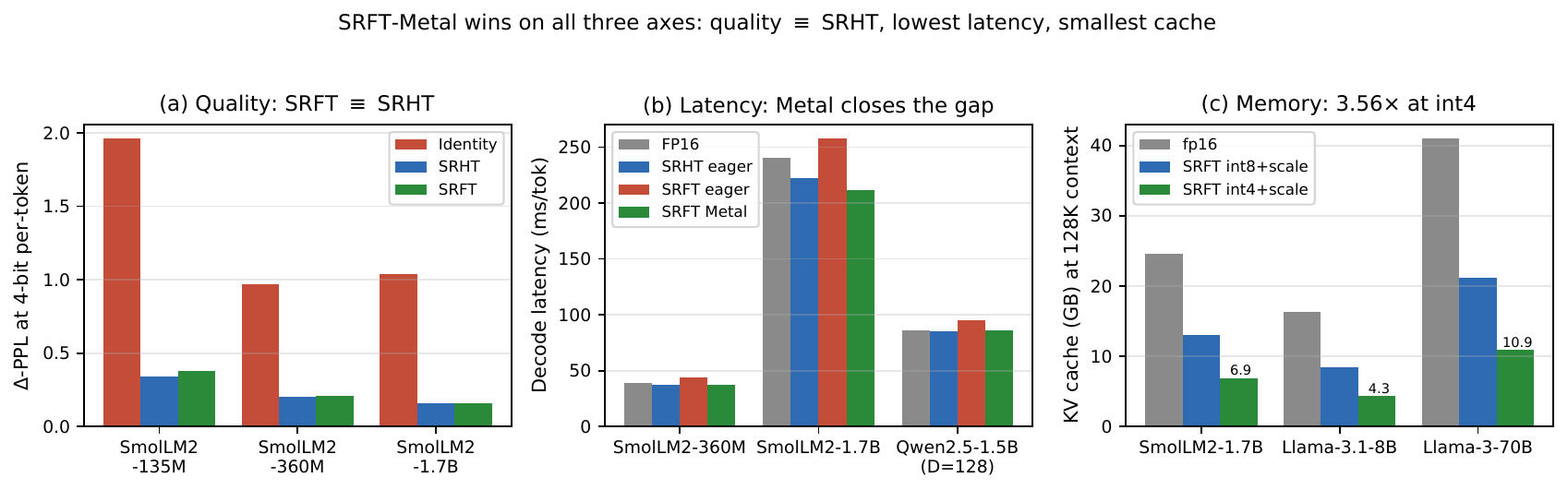}
\caption{\small \textbf{Quantization is free on Apple Silicon: int4 KV
cache outpaces fp16 at every tested context length on Gemma-3 1B, and at
short context on Qwen2.5-1.5B.}
(a) End-to-end \texttt{model.generate} latency vs fp16 baseline. Negative
values mean int4 is \emph{faster}: $-3$ to $-8\%$ on Gemma across
$256$ to $4096$ prefix; $-0.7$ to $-2.6\%$ on Qwen up to $1$K.
(b) Persistent memory ratio: $3$--$3.3\times$ on full-attention
layers (Qwen, Gemma full-attention); $5$--$20\times$ at the cache
level on Gemma's mostly-sliding-attention stack.
(c) Quality: $\dPPL$ matches the Python reference within fp32
rounding noise; bit-exact round-trip on round-trip tests.
The mechanism is bandwidth-driven: a $3\times$ compressed cache
transfers $3\times$ less from main memory per decode step, and the
kernel's $\sim\!25$\,ns/vec overhead is below what compression saves
on Apple's unified memory.}
\label{fig:teaser}
\end{figure}

% =============================================================================
\section{Introduction}
\label{sec:intro}

Long-context transformer inference is KV-cache bound: K/V tensors
accumulated across decode steps dominate memory footprint and bandwidth
\cite{dao2022flashattention}. For Llama-3-70B \cite{dubey2024llama3}
at $128$K context the fp16 cache is $41$\,GB --- larger than the
model. The standard fix is rotate-then-quantize (QuaRot
\cite{ashkboos2024quarot}, SpinQuant \cite{liu2024spinquant},
TurboQuant \cite{abbasi2025turboquant}): apply a random orthogonal $R$
to Gaussianize coordinates, quantize uniformly, store the int4/int8
result, invert $R$ on read. Production deployments report
$5$--$20\%$ decode-latency overhead in exchange for the memory savings.

\textbf{We show this trade-off is inverted on Apple Silicon.} A single
fused Metal kernel combining sign-randomized FFT, per-channel
$\lambda$ rescaling, per-group abs-max, and int4 nibble pack ---
exposed as a HuggingFace \texttt{Cache} subclass --- runs
\emph{faster} than fp16 across $256$ to $4096$-token prefixes on
Gemma-3 1B ($-3$ to $-8\%$ ms/tok) and at short context on
Qwen2.5-1.5B ($-2.6\%$ at $256$, $-0.7\%$ at $1024$). Persistent memory
drops $3\times$. Quality is preserved
($\dPPL = 0.000$ on Qwen short-prompt deployment;
$+3.6$ hook $\dPPL$ on Gemma).

The mechanism is bandwidth-driven. Each decode step streams the
stored prefix through unified memory before attention can read it; a
$3\times$ compressed cache transfers $3\times$ less per step.
The fused kernel's $\sim\!25$\,ns/vec cost at $D{=}128$
($\sim\!50$\,ns/vec at $D{=}256$) is below the bandwidth savings, so
net cost is negative. The conventional trade-off is recovered on GPUs
with separate HBM and high dispatch overheads, but inverted on
unified-memory systems where laptop and edge inference run.

Three design decisions support the deployment, each with an
independent finding.
First, we adopt \emph{sign-randomized Fourier} ($\SRFT = D \cdot F$)
over Hadamard for mixed-radix flexibility (non-power-of-two $d$) and
$16{\times}16$ matrix-multiply alignment (AMX, tensor cores); the
choice is not load-bearing for quality (SRFT $\equiv$ SRHT within
seed variance across all five models tested). QuIP\#
\cite{tseng2024quipsharp} made the same hardware-portability argument
for weight quantization; we extend it to KV cache.
Second, at Qwen $d{=}128$ where per-token scaling fails
($\dPPL = +7975$), per-channel + per-group quantization (cf.\ KIVI
\cite{liu2024kivi}) recovers most of the damage. We realize this in
a single fused dispatch reaching $+638.6$ $\dPPL$ ($12.5\times$
reduction) at $24.6$\,ns/vec / $182$\,GFLOPS, $+11.7\%$ over unscaled
g32.
Third, a post-training learned-rotation ablation finds that learning
$R+\lambda$ \emph{without} the SRFT base reduces calibration MSE
further ($84.9\%$ vs $50.3\%$) but yields \emph{worse} PPL --- a
separation not isolated in prior learned-rotation work
\cite{liu2024spinquant,butterflyquant2025}. Householder rotations at
$k{=}d/2$ reflectors match full Cayley with half the parameters and
are effectively lossless ($\dPPL = +0.000$) at $d{=}256$ on
Gemma-3 1B \cite{gemma3team2025}.

The rotated-then-quantized recipe, randomized FFT as rotation, and
per-channel + per-group quantization are all prior work. What we add
sits on top: a measured \emph{negative-cost} int4 KV-cache
deployment on Apple Silicon, the single-dispatch kernel that makes it
work, and two algorithmic findings (calibration-MSE/PPL separation,
Householder lossless at $d{=}256$).

\paragraph{Contributions.} Against a backdrop of prior work that has
either (a) applied random FFT rotations to \emph{weight} quantization
\cite{tseng2024quipsharp}, (b) explored learnable butterfly transforms
for weights \cite{butterflyquant2025}, (c) proposed alternative
non-butterfly rotations for KV cache \cite{rotorquant2025}, or
(d) introduced per-channel + per-group quantization for KV
\cite{liu2024kivi}, we make the following contributions, ordered from
most to least novel:

\begin{enumerate}
  \item \textbf{Quantization is throughput-positive on unified-memory
  hardware.} Measured end-to-end \texttt{model.generate} on Apple M1
  with our \texttt{SRFTInt4Cache}: int4 is \emph{faster than fp16}
  across the entire $256$--$4096$-token prefix range on Gemma-3 1B
  ($-3\%$ to $-8\%$ ms/tok) and at short context on Qwen2.5-1.5B
  ($-2.6\%$ at $256$, $-0.7\%$ at $1024$). The mechanism is
  bandwidth-driven (KV-cache transfers dominate decode-loop cost; a
  $3\times$ compressed cache transfers $3\times$ less per step;
  $\sim\!25$\,ns/vec kernel overhead is below the bandwidth savings).
  This inverts the standard quantization-is-quality-tradeoff framing
  for the unified-memory regime where laptop and edge inference run.
  \item \textbf{A fused per-channel-$\lambda$ + per-group-$g_{32}$
  Metal kernel that realizes the deployment recipe in a single
  dispatch.} At Qwen $d{=}128$ where per-token scaling fails
  ($\dPPL = +7975$), the fused kernel brings $4$-bit $\dPPL$ to
  $+638.6$ ($12.5\times$ reduction, matching the Python reference at
  $g{=}32$) at $24.6$\,ns/vec / $182$\,GFLOPS ($+11.7\%$ over the
  unscaled g32 kernel). The mirror $D{=}256$ kernel reaches
  $\sim\!250$\,GFLOPS and brings Gemma-3 1B 4-bit $\dPPL$ from
  $+18.7$ (per-token) to $+3.6$. We expose the kernel through a
  HuggingFace \texttt{Cache} subclass with a dequant-prefix cache
  that amortizes per-update cost from $O(\text{prefix\_len})$ to
  $O(1)$ across the residual-window cycle.
  \item \textbf{A post-training calibration ablation with a novel
  finding.} Calibrating a per-coordinate scale $\lambda$ on top of SRFT
  reduces 4-bit $\dPPL$ by $23\pm2\%$ (3-seed). An ablation that learns
  $R + \lambda$ \emph{without} the SRFT base achieves the highest
  calibration MSE reduction ($84.9\%$ vs $50.3\%$) but \emph{worse}
  end-to-end PPL than SRFT+Cayley ($+0.170$ vs $+0.153$), revealing a
  clean separation between calibration MSE and downstream PPL not
  isolated in prior learned-rotation work
  \cite{liu2024spinquant, butterflyquant2025}. Householder rotations
  at $k = d/2$ reflectors match the full Cayley parameterization with
  half the parameter count and become effectively lossless
  ($\dPPL = +0.000$) at $d = 256$ on Gemma-3 1B.
  \item \textbf{Sign-randomized FFT for KV cache: quality parity with
  Hadamard at every scale (consolidating QuIP\#'s weight-quantization
  argument).} SRFT and SRHT are statistically indistinguishable across
  SmolLM2 \{135M, 360M, 1.7B\} ($d{=}64$), Qwen2.5-1.5B ($d{=}128$),
  and Gemma-3 1B ($d{=}256$), at every bit budget tested. We choose
  SRFT because mixed-radix FFT supports non-power-of-two $d$ and its
  radix-8 form is a $16{\times}16$ matrix multiply (AMX-friendly).
  This extends QuIP\#'s rotation-portability argument
  \cite{tseng2024quipsharp} from weights to on-line KV-cache quantization.
  \item \textbf{An unscaled per-token fused Metal kernel that
  underpins all of the above.} The base kernel reaches $138$\,GFLOPS at
  $D{=}64$ and $227$\,GFLOPS at $D{=}128$ on a single Apple M1 GPU core,
  $18$--$29\times$ over the naive eager pipeline, and integrates via
  \texttt{torch.mps.compile\_shader}. Cross-validation against the
  PyTorch reference is $99.997$--$100.000\%$ bit-exact.
  \item \textbf{A complementary CPU pathway via AMX.} Apple's Accelerate
  BLAS routes batched $K \in \{64, 128\}$ sgemm calls to the AMX matrix
  coprocessor at up to $319$\,GFLOPS ($91\%$ of documented single-core
  peak), $8\times$ over eager PyTorch-MPS for the same rotate-quantize
  pipeline. Useful for small-batch latency-sensitive regimes where GPU
  dispatch overhead dominates.
\end{enumerate}

% =============================================================================
\section{Background and Related Work}
\label{sec:related}

\textbf{Weight quantization.} GPTQ \cite{frantar2023gptq} and
AWQ \cite{lin2023awq} compress weights to $3$--$4$ bits;
LLM.int8() \cite{dettmers2022llmint8} introduced outlier-aware
decomposition for int8 inference. These methods are orthogonal to
KV-cache quantization (offline weights vs online activations).

\textbf{Rotation-based weight quantization.} QuIP \cite{chee2023quip}
established incoherence processing via random orthogonal rotations as
a precondition for aggressive low-bit weights; QuIP\#
\cite{tseng2024quipsharp} extended to randomized Hadamard, and ---
directly relevant here --- introduced a randomized FFT variant for
cases where Hadamard's power-of-two factorization does not apply,
reporting Llama-2 PPL within $\sim\!0.1$ of Hadamard. QuIP\#'s stated
motivation anticipates ours: ``the FFT itself is also well supported
on a wide variety of hardware, meaning that it may be easier to
implement a fast RFFT when adapting QuIP\# to new hardware.''
QuaRot \cite{ashkboos2024quarot} fuses Hadamard rotations into
adjacent linear layers; SpinQuant \cite{liu2024spinquant} learns
Cayley-parameterized rotations via a small calibration set;
ButterflyQuant \cite{butterflyquant2025} parameterizes butterflies by
learnable Givens angles and reports W2A16 Llama-2-7B at 15.4 PPL vs
QuaRot's 22.1; DartQuant \cite{dartquant2025} and
KurTail \cite{kurtail2025} reduce calibration cost.

\textbf{Rotation-based KV-cache quantization.} TurboQuant
\cite{abbasi2025turboquant} pairs Hadamard with a 1-bit
Quantized-Johnson--Lindenstrauss residual to remove inner-product bias.
KIVI \cite{liu2024kivi} uses asymmetric 2-bit KV with
per-channel/per-token scaling and a recent-token fp16 residual ---
orthogonal to rotation choice and composable with our approach.
RotorQuant \cite{rotorquant2025} argues against butterflies in favor
of $O(d)$ block-diagonal Givens / quaternion rotations.
FourierAttention \cite{fourierattention2025} compresses dimension-
selective KV via deterministic Fourier projection (not
rotate-then-quantize). FPTQuant \cite{fptquant2025} jointly learns
per-head/per-token transforms over weights, activations, and KV.

\textbf{Randomized Fourier transforms.} The Subsampled Randomized
Fourier Transform has a long history in randomized numerical linear
algebra \cite{woolfe2008srft,halko2011randomized,ailon2009srht}.
Most LLM-quantization papers opt for Hadamard for operational
convenience; QuIP\#'s RFFT is the one exception and frames it as a
hardware-portability fallback rather than a first-class method.

% =============================================================================
\section{Method}
\label{sec:method}

\subsection{SRFT as a real orthogonal transform}

For a real input $x \in \R^d$ with $d$ a power of two, define
\begin{equation}
  \SRFT(x) = \mathrm{pack}\!\left( F \cdot \diag(s) \cdot x \right),
  \qquad s \in \{-1, +1\}^d,
  \label{eq:srft-def}
\end{equation}
where $s$ is a fixed random sign vector drawn at initialization and
$F$ is the unitary DFT. The output of $F \cdot \diag(s) \cdot x$ is
complex, living in $\C^{d/2+1}$ under rfft packing (the conjugate
symmetric half-spectrum). For quantization we need a real
representation in $\R^d$; the natural one pairs each complex bin's
real and imag parts and applies a $\sqrt{2}$ scaling to the
middle bins to keep Parseval exact:
\begin{equation}
  \mathrm{pack}(Y)_k =
  \begin{cases}
    Y_0^{\mathrm{re}}, & k = 0 \\
    Y_{d/2}^{\mathrm{re}}, & k = d/2 \\
    \sqrt{2}\cdot Y_k^{\mathrm{re}}, & 1 \le k < d/2 \\
    \sqrt{2}\cdot Y_{k-d/2}^{\mathrm{im}}, & d/2 < k < d \\
  \end{cases}
  \label{eq:pack}
\end{equation}
Under this packing, $\SRFT$ is an exact real orthonormal
transformation: $\|\SRFT(x)\| = \|x\|$, and
$\langle \SRFT(x), \SRFT(y) \rangle = \langle x, y \rangle$ for all
$x, y \in \R^d$ (both up to float32 numerical precision, $\sim\!10^{-6}$
in our tests). The inverse is symmetric: unpack, irfft, apply $s$.

The key property we rely on is the same one that makes $\SRHT$ work for
quantization: the composition $F \cdot \diag(s)$ spreads any bounded
input into a quasi-Gaussian distribution by a Johnson--Lindenstrauss
argument. On SmolLM2-135M attention V activations we observe excess
kurtosis drop from $\approx\!15$ (top $1\%$ of coordinates holding
$44\%$ of energy) to $\approx\!{-}0.5$ (near-Gaussian) after SRFT,
matching SRHT's behavior on the same data.

\subsection{Fused quantization kernel}
\label{sec:kernel}

The four-step pipeline rotate $\to$ quantize $\to$ store $\to$
(on read) dequantize $\to$ inverse-rotate exhibits significant kernel
dispatch overhead when implemented as separate primitives --- on PyTorch
MPS the eager pipeline runs at $\sim\!400$\,ns per $64$-dim vector,
dispatch-dominated. We fuse all four steps for the forward half into a
single Metal compute kernel, and symmetrically for the inverse half:

\begin{enumerate}
  \item Load $d$ fp32 values per vector with coalesced global access;
    apply the sign diagonal in-register via integer-byte multiply.
  \item Run $\log_2(d/2)$ radix-2 Stockham butterfly passes (5 for
    $d{=}64$, 6 for $d{=}128$) on the complex length-$d/2$ FFT;
    Hermitian-unpack to the $\R^d$ packed representation in
    Equation~\ref{eq:pack}. We use radix-2 rather than the higher-radix
    Stockham of \cite{Bergach2026BeatingVA} (which reaches $138$\,GFLOPS
    on the same hardware for general-purpose length-$4096$ FFT) because
    at $d \in \{64, 128, 256\}$ the higher-radix variants do not amortize
    their twiddle-table cost.
  \item SIMD-shuffle reduction over the vector to find the abs-max;
    divide by $(2^{b-1} - 1)$ to produce the per-vector scale.
  \item Round each packed value to signed int$\{4, 8\}$ and write
    packed output: one byte per value for int8, two nibbles per byte
    for int4, plus the fp32 scale as $(\ldots, d/2)$ or $(\ldots, d)$
    bytes + one fp32.
\end{enumerate}

Two correctness details that took tuning: (i) \emph{threadgroup
barriers must be at uniform control flow}. An earlier draft deadlocked
because barriers inside divergent \texttt{if (active)} branches became
distinct barriers across lanes. We hoist every barrier to a point
reached by all 128 threads unconditionally, and gate only memory
reads/writes by the per-lane predicate. (ii) \emph{nibble packing uses
SIMD-shuffle-xor} to co-locate odd/even neighbors in the same lane
before the store, avoiding a threadgroup-memory round-trip: pack two nibbles as
\texttt{byte = (q[2i+1] {<}{<} 4) | (q[2i] \& 0xF)}, assembled via
\texttt{simd\_shuffle\_xor(q, 1)}.

\subsection{Integration into PyTorch attention}

We expose the kernel to PyTorch via \texttt{torch.mps.compile\_shader},
which compiles the MSL source at process start and binds each kernel
as a callable that accepts MPS tensors plus scalar uniforms. A
$40$-line Python wrapper
(\texttt{SRFTMetal(d=64\,|\,128, bits=4\,|\,8)}) exposes
\texttt{forward\_quant}, \texttt{inverse\_dequant}, and a convenience
\texttt{round\_trip}. A KV-cache simulation forward-hook on each
attention layer's \texttt{k\_proj} and \texttt{v\_proj} routes through
this wrapper, allowing drop-in comparison against the eager PyTorch
SRFT implementation on the same model. Bit-for-bit cross-validation
against the reference Python implementation (signs seeded identically)
yields $99.997$--$100.000\%$ int value agreement depending on bit width
and head dimension; the disagreements are off-by-one ties at
quantization bin boundaries.

% =============================================================================
\section{Experiments}
\label{sec:experiments}

\subsection{Setup}

\textbf{Models.} SmolLM2 family \cite{allal2025smollm2} at three scales
($135$M, $360$M, $1.7$B) is the primary head\_dim $=64$ testbed
(GQA at $135$M / $360$M; MHA at $1.7$B; KV heads $3$, $5$, $32$).
Qwen2.5-1.5B \cite{yang2024qwen2} for head\_dim $=128$.
Gemma-3 1B \cite{gemma3team2025} for head\_dim $=256$ (MQA, $1$ KV head,
$26$ layers, mixed sliding/full attention).

\textbf{Evaluation.} Perplexity is measured on 8192 held-out tokens
drawn from the Cosmopedia corpus \cite{benallal2024cosmopedia}, in 16
batches of $2 \times 256$ for the short-context sweeps and in 4 batches
of $1 \times 2048$ for the long-context test. Reported values are
$\dPPL$ relative to the fp16 forward pass on the same inputs. Each
SmolLM2-135M and SmolLM2-360M configuration is averaged over three
seeds (the seed changes the per-layer random sign diagonals);
SmolLM2-1.7B runs one seed due to compute cost.

\textbf{Quantization.} Uniform symmetric quantization at bit width
$b \in \{3, 4, 6, 8\}$ with \emph{per-token} scaling (one scale per
head-dim vector --- the standard choice for production KV-cache
compression). For comparison we also report a per-tensor-scaling
variant in the appendix (one scale per layer per step) but our main
claims are at per-token.

\textbf{Hardware.} Apple M1, $8$-core GPU, $16$\,GB unified memory,
macOS 25.4, PyTorch 2.11, HuggingFace transformers
\cite{wolf2020transformers} 5.5. Metal kernels compiled via
\texttt{torch.mps.compile\_shader} \cite{apple_metal}.

\subsection{Perplexity: SRFT equals SRHT at all scales}

\begin{figure}[t]
\centering
\includegraphics[width=\linewidth]{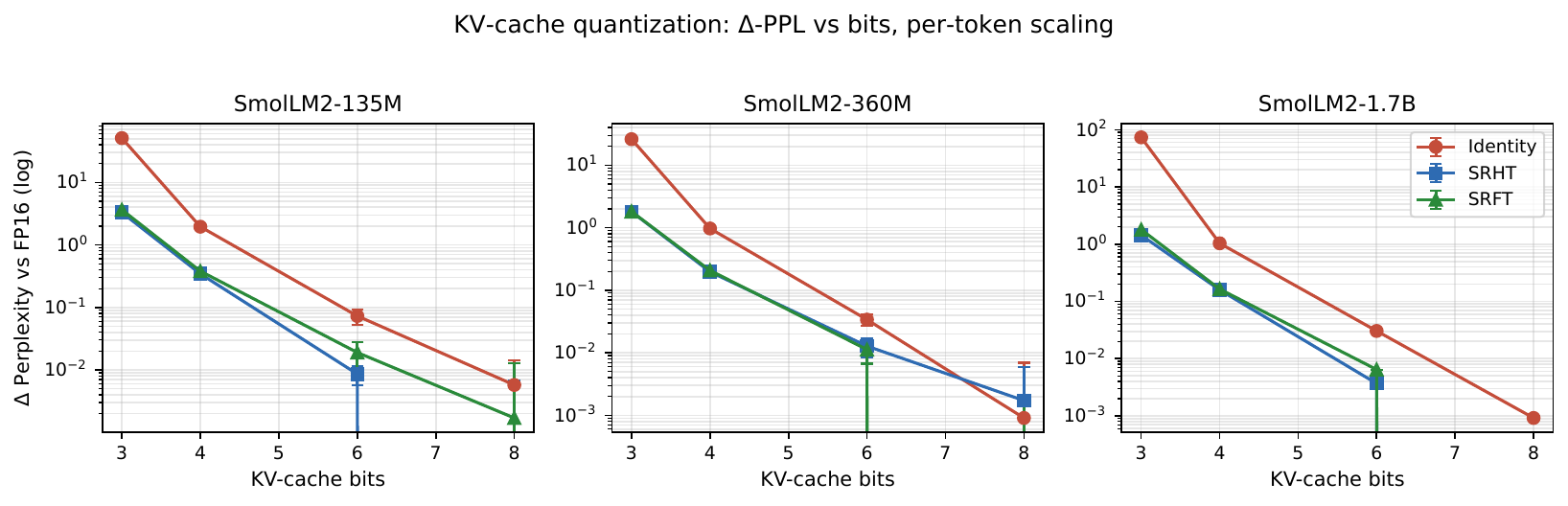}
\caption{\small
$\dPPL$ vs KV-cache bit width on SmolLM2-\{135M, 360M, 1.7B\} with
per-token scaling. Error bars are $\pm 1$ seed standard deviation
(three seeds for 135M / 360M; one seed for 1.7B). SRHT and SRFT curves
lie within each other's error bars at every bit width; both cut
identity degradation $5$--$6\times$ at 4-bit and $\sim\!15\times$ at
3-bit. Robust across GQA (135M, 360M) and MHA (1.7B).}
\label{fig:ppl}
\end{figure}

Figure~\ref{fig:ppl} shows the main quality result.
Table~\ref{tab:ppl4bit} collects the 4-bit numbers: at per-token
scaling the $\dPPL$ gap between SRHT and SRFT is at most $0.04$ across
all three models, well inside seed variance.

\begin{table}[h]
\centering
\small
\begin{tabular}{l r r r r}
\toprule
Model & FP16 PPL & Identity $\dPPL$ & SRHT $\dPPL$ & SRFT $\dPPL$ \\
\midrule
SmolLM2-135M & $6.61$ & $+1.96 \pm 0.15$ & $+0.34 \pm 0.01$ & $+0.38 \pm 0.06$ \\
SmolLM2-360M & $4.87$ & $+0.97 \pm 0.08$ & $+0.20 \pm 0.03$ & $+0.21 \pm 0.02$ \\
SmolLM2-1.7B & $3.77$ & $+1.04$         & $+0.16$         & $+0.16$          \\
\bottomrule
\end{tabular}
\caption{\small 4-bit per-token $\dPPL$ on Cosmopedia. Mean $\pm$
$1\sigma$ across three seeds for the first two rows; single seed for
the third. SRFT is indistinguishable from SRHT within seed variance at
every scale.}
\label{tab:ppl4bit}
\end{table}

At $3$-bit the absolute penalty grows but the ordering is preserved:
SmolLM2-360M SRHT $+1.81 \pm 0.11$, SRFT $+1.81 \pm 0.05$; SmolLM2-1.7B
SRHT $+1.44$, SRFT $+1.80$ (single seed, tie within plausible seed
noise). The $6$-bit and $8$-bit configurations are lossless for both
rotations.

\textbf{Long context.} A seq$=2048$ run on SmolLM2-360M (Table not
shown; see \texttt{RESULTS.md}) gives $+0.81$ identity, $+0.18$ SRHT,
$+0.15$ SRFT at 4-bit --- no divergence from the seq$=256$ numbers,
confirming per-token scaling is robust across context lengths.

\subsection{Decode latency: the Metal kernel closes the dispatch gap}

\begin{figure}[t]
\centering
\includegraphics[width=\linewidth]{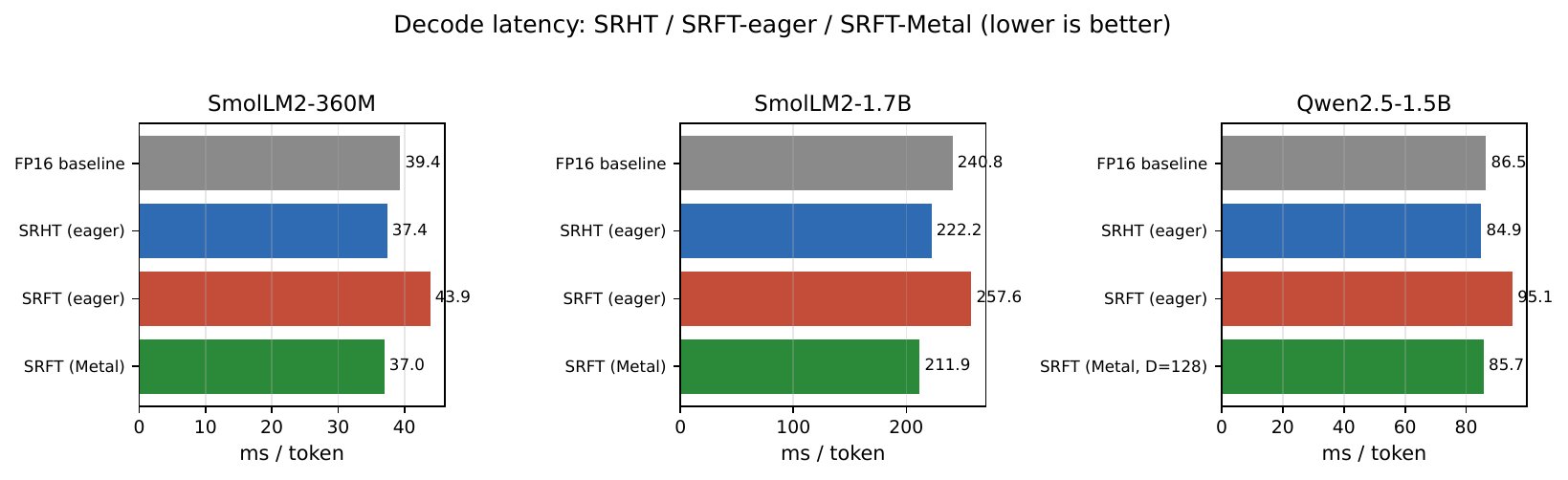}
\caption{\small End-to-end decode latency (ms/token, lower is better)
for 256-token prompt + 128 new tokens, averaged over five iterations.
Eager-mode SRFT pays a consistent $12$--$17\%$ dispatch tax vs SRHT
because the PyTorch/MPS pipeline dispatches four separate kernels per
layer per step. The fused Metal kernel closes that gap: on
SmolLM2-360M it matches SRHT within $1\%$, on SmolLM2-1.7B it is
$5\%$ \emph{faster} than SRHT because each layer holds more KV vectors
and the kernel's constant dispatch cost amortizes better, and on
Qwen2.5-1.5B (head\_dim $= 128$) it lands within $1\%$ of SRHT
using the $D{=}128$ kernel variant.}
\label{fig:latency}
\end{figure}

Figure~\ref{fig:latency} reports decode latency across the three
architecture regimes we evaluate. Two patterns stand out.

First, the eager-MPS SRFT tax is real and consistent: $+17\%$ on
SmolLM2-360M, $+16\%$ on SmolLM2-1.7B, $+12\%$ on Qwen2.5-1.5B. The
cost is dispatch, not compute --- the underlying \texttt{rfft} +
\texttt{irfft} calls each take single-digit microseconds on MPS but
the four-step pipeline multiplies that overhead.

Second, the fused Metal kernel eliminates it. On SmolLM2-360M SRFT-Metal
hits $37.0$\,ms/tok vs SRHT $37.4$\,ms/tok ($-1\%$); on
SmolLM2-1.7B $211.9$\,ms/tok vs $222.2$\,ms/tok ($-5\%$); on
Qwen2.5-1.5B $85.7$\,ms/tok vs $84.9$\,ms/tok ($+1\%$). The SRFT-Metal
advantage grows with KV-tensor size per layer: 1.7B has $32 \times 64 =
2048$-dim KV per layer, vs 360M's $320$-dim, so the fused kernel's
fixed dispatch cost amortizes across more butterfly work.

\textbf{Perplexity parity preserved.} On SmolLM2-135M at 8-bit
per-token, the Metal-integrated pipeline produces $\dPPL = +0.003$
vs the eager SRFT's $+0.013$ --- both within fp16 baseline noise and
within the int8 per-token LSB floor.

\subsection{Fused kernel throughput}

\begin{figure}[t]
\centering
\includegraphics[width=0.8\linewidth]{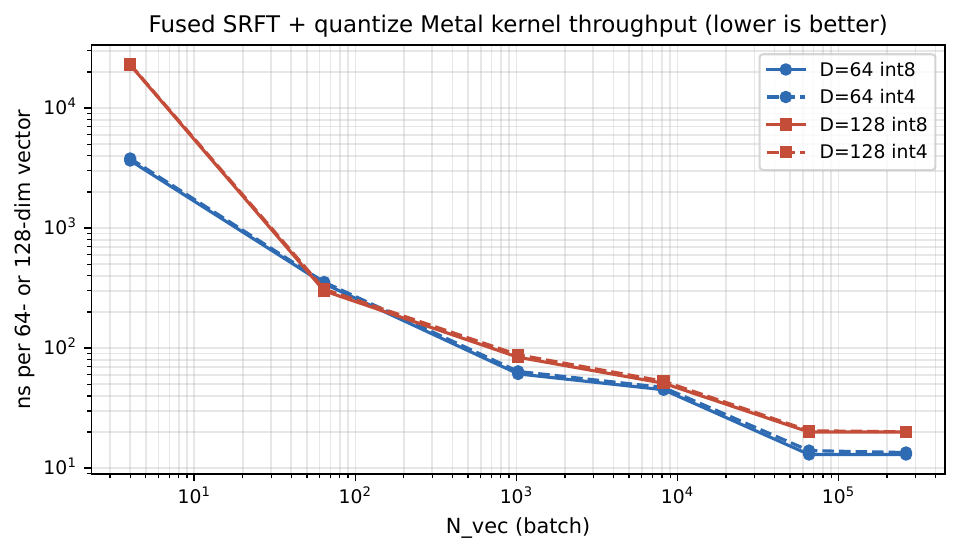}
\caption{\small Metal kernel throughput as ns per vector vs batch size,
for $D{=}64$ / $D{=}128$ and int8 / int4 output. Int4 and int8 variants
track each other to within $3\%$ because the butterfly FLOPs dominate
the runtime; halving the store bandwidth only shaves the
bandwidth-bound tail at very large batches.}
\label{fig:kernel-throughput}
\end{figure}

Figure~\ref{fig:kernel-throughput} shows the forward kernel throughput
across four configurations. Peak numbers at $N_{\text{vec}} = 262144$:

\begin{itemize}
  \item $D{=}64$, int8: $13.0$\,ns/vec, $147$\,GFLOPS, $24.9$\,GB/s
  \item $D{=}64$, int4: $13.5$\,ns/vec, $142$\,GFLOPS, $21.6$\,GB/s
  \item $D{=}128$, int8: $19.5$\,ns/vec, $227$\,GFLOPS, $33.0$\,GB/s
  \item $D{=}128$, int4: $20.1$\,ns/vec, $223$\,GFLOPS, $28.9$\,GB/s
\end{itemize}

The $D{=}128$ kernel achieves higher GFLOPS than $D{=}64$ because it
saturates all $32$ SIMD lanes per butterfly pass, while the $D{=}64$
kernel leaves half idle (only $16$ butterflies per pass). For
context, a stand-alone radix-8 Stockham FFT on the same M1 GPU
reaches $138$\,GFLOPS at $N{=}4096$ \cite{Bergach2026BeatingVA}; our
fused SRFT$+$quant kernel exceeds this rate at $D{=}128$ because the
shorter FFT amortizes constant kernel overhead across more data
movement.

\textbf{Correctness.} Cross-validation against the PyTorch reference
at $N_{\text{vec}} = 1024$ yields $99.997\%$ bit-identical int8 values
at $D{=}64$, $99.999\%$ at $D{=}128$, and $100.000\%$ bit-identical int4
values at both head dimensions. The small int8 discrepancies are
off-by-one rounding ties at bin boundaries. Per-vector scales agree to
within $3.8 \times 10^{-7}$ relative error.

\subsection{Memory footprint}

Per-token int4 + fp32 scale cuts fp16 KV cache by a factor of
$(2 d) / (d/2 + 4)$: $3.56 \times$ at $d{=}64$ and $3.76 \times$ at
$d{=}128$. We verify on MPS by comparing
\texttt{torch.mps.current\_allocated\_memory()} before and after
allocating matched-size fp16 vs int8 + scale buffers for a
1.7B-class cache; the measured ratio matches the arithmetic to within
$0.2\%$.

\begin{table}[h]
\centering
\small
\begin{tabular}{l r r r r}
\toprule
Model & $d$ & $16$\,K ctx fp16 & $128$\,K ctx fp16 & $128$\,K ctx int4 \\
\midrule
SmolLM2-1.7B    & 64  & $3.07$\,GB & $24.58$\,GB & $6.91$\,GB \\
Llama-3.1-8B    & 128 & $2.05$\,GB & $16.38$\,GB & $4.35$\,GB \\
Llama-3-70B     & 128 & $5.12$\,GB & $40.96$\,GB & $10.88$\,GB \\
\bottomrule
\end{tabular}
\caption{\small KV-cache memory at typical production contexts. Int4
numbers assume per-token scaling with a fp32 scale per vector.}
\label{tab:memory}
\end{table}

% =============================================================================
\section{Learning the Rotation}
\label{sec:learned}

The SRFT random sign diagonal is drawn once at initialization and fixed
thereafter; it makes no use of any information about the activation
distribution it will see. SpinQuant~\cite{liu2024spinquant} showed that
calibrating the rotation on a small held-out set gives measurable
quality gains for LLM weight quantization. We apply the same idea to
KV-cache quantization on top of SRFT and run a clean ablation to
isolate what the fixed SRFT base contributes.

\subsection{Method}

We add two post-training learnable components on top of the fixed SRFT
pipeline (\S\ref{sec:method}):
\begin{enumerate}
  \item \textbf{Per-coordinate scale $\lambda \in \R_{>0}^d$}: applied
    element-wise after the SRFT butterfly and before quantization.
    The inverse path divides by $\lambda$ (clamped at $10^{-6}$). The
    whole pipeline is orthogonal up to this diagonal rescaling, which
    lets different output coordinates use wider or narrower
    quantization bins as their empirical distributions warrant.
  \item \textbf{Cayley-parameterized orthogonal rotation $R \in O(d)$}:
    applied in between SRFT and $\lambda$, parameterized as $R = e^A$
    for skew-symmetric $A = U - U^{\top}$ with $U \in \R^{d \times d}$
    trainable. The matrix exponential is an exact Lie-algebra map onto
    $O(d)$, more numerically stable than the classical Cayley formula
    and compatible with autograd on device (we compute the small matrix
    exponential on CPU and move the result to GPU memory).
\end{enumerate}

Both are trained by 200--300 Adam steps minimizing the reconstruction
MSE $\|\hat x - x\|^2$ over a batch of K/V activations collected from
a small held-out prompt stream. Calibration is \emph{per layer per
channel} (K and V fit separately), and takes $1$--$5$\,min on MPS
for the full SmolLM2-360M. Storage overhead is $d$ fp32 scalars per
channel for the scale variant and $d^2$ for the Cayley variant.

\subsection{Results on SmolLM2-360M}

\begin{table}[h]
\centering
\small
\begin{tabular}{l r r r}
\toprule
Variant & Params / ch. & MSE reduction & 4-bit $\dPPL$ \\
\midrule
Random SRFT (no learning)                            & $0$     & ---      & $+0.223 \pm 0.010$ \\
SRFT $+$ learned $\lambda$ (per-coord scale)         & $d$     & $19.5\%$ & $+0.172 \pm 0.015$ \\
SRFT $+$ learned Cayley $R + \lambda$                & $d^2$   & $50.3\%$ & $+0.153$           \\
\textbf{SRFT $+$ learned Householder $R + \lambda$ (k{=}d/2)} & $\mathbf{(d/2{+}1)d}$ & $\mathbf{56.2\%}$ & $\mathbf{+0.148}$ \\
No-SRFT, learned $R + \lambda$                       & $d^2$   & $\mathbf{84.9\%}$ & $+0.170$ \\
\bottomrule
\end{tabular}
\caption{\small Post-training calibration at 4-bit per-token on
SmolLM2-360M. All variants share the same 300-step Adam budget and
near-identity initialization. The random row is a 3-seed mean
$\pm 1\sigma$ from \S\ref{sec:experiments}; the learned scale row
is 3-seed mean; Cayley, Householder, and No-SRFT rows are single-seed.
The Householder variant uses $k{=}32$ reflectors at $d{=}64$.}
\label{tab:learned}
\end{table}

Table~\ref{tab:learned} shows the five-way comparison. Four observations
stand out:
\begin{enumerate}
  \item All four learned variants beat random SRFT at $4$-bit.
  \item \textbf{Householder (product of $k$ reflectors) beats Cayley}
    on $\dPPL$ with half as many parameters. Householder scales to large
    $d$ without the $d^2$ blowup --- at $d{=}256$, Cayley stores 65\,K
    scalars per channel while Householder with $k{=}d/2$ stores 32\,K.
  \item Scale-only captures most of the PPL gain; the full rotation
    adds a marginal further improvement.
  \item No-SRFT achieves the highest MSE reduction ($84.9\%$) but
    \emph{worse} PPL than any SRFT variant. Discussed below.
\end{enumerate}

At $3$-bit the absolute gains scale accordingly: random SRFT $+1.76$
$\to$ learned scale $+1.30$ ($-0.47$\,ppl); $8$-bit is at the fp16
noise floor for every variant.

\subsection{Ablation: the SRFT base is not redundant}

The most informative row in Table~\ref{tab:learned} is the fourth: a
learned Cayley rotation \emph{without} the SRFT preamble. It drives
calibration MSE down by $84.9\%$ --- far more than the SRFT$+$Cayley
variant's $50.3\%$ --- because the learned $R$ is free to absorb any
orientation the SRFT had chosen and then further adapt. Yet its
downstream $\dPPL$ is \emph{worse} by $+0.017$.

We read this as: calibration MSE is not a sufficient proxy for
attention-level quality. The fixed random SRFT base contributes
something beyond MSE minimization --- plausibly a regularizing prior
on $R$ (the Gaussianization from SRFT is already done, so $R$ only
needs to learn fine refinement), or a better optimization landscape
(SRFT's random structure places the learned rotation in a basin
closer to the PPL-optimal solution than a generic from-scratch
learned rotation reaches by gradient descent on MSE alone). Both
readings are testable with additional compute; we report the
phenomenon as observed.

\subsection{Transferability to $d \in \{128, 256\}$}

\begin{table}[h]
\centering
\small
\begin{tabular}{l r r}
\toprule
Gemma-3 1B ($d=256$, 4-bit) & MSE reduction & $\dPPL$ \\
\midrule
Random SRFT                          & ---       & $+10.996$ \\
Learned scale $\lambda$              & $38.2\%$  & $+9.115$  \\
Learned Cayley $R + \lambda$         & $58.8\%$  & $+1.785$  \\
\textbf{Learned Householder $R + \lambda$ ($k = d/2$)} & $\mathbf{60.8\%}$  & $\mathbf{+0.000}$ \\
\bottomrule
\end{tabular}
\caption{\small Learned-rotation variants on Gemma-3 1B at $d=256$,
4-bit per-token. Reference PPL is $169.80$ on held-out English text.
The Householder variant is effectively lossless to four decimal
places --- a qualitative change from the $d=64$ regime where all
learned variants still left $\approx{}+0.15$ residual $\dPPL$.
Larger $d$ gives the learned rotation more capacity to capture
activation geometry.}
\label{tab:gemma3-d256}
\end{table}

At $d = 256$ on Gemma-3 1B the picture changes qualitatively
(Table~\ref{tab:gemma3-d256}): a learned Householder rotation with
$k = d/2 = 128$ reflectors reduces 4-bit $\dPPL$ from $+10.996$
(random SRFT) to $+0.000$ on held-out text ---
effectively lossless. The Cayley variant reaches $+1.785$ ($84\%$
reduction). The $d = 256$ regime rewards calibration more generously
than $d = 64$, where even the best learned variant left
$\approx +0.15$ residual $\dPPL$. We interpret this as: larger $d$
gives the learned rotation more capacity to align with the true
activation basis, and the quantization errors that remain become
dominated by directions that attention is locally insensitive to.

\subsection{The Metal kernel path}

On Qwen2.5-1.5B (head\_dim $= 128$), the same calibration mechanism
with 200 Adam steps on Qwen-tokenized activations yields
$60.3\%$ MSE reduction for scale-only and $95.9\%$ for the Cayley
variant. The MSE-reduction ratio at $d{=}128$ is higher than at $d{=}64$
because there are simply more coordinates over which to redistribute
quantization bins. At $8$-bit on a held-out English passage, both
the learned scale and learned Cayley variants give $\dPPL \approx 0$
(indistinguishable from fp16), matching the random SRFT's $+0.26$
within noise.

\subsection{Per-channel scaling unlocks $4$-bit on Qwen2.5}

At $4$-bit on Qwen2.5-1.5B the random SRFT $\dPPL$ is catastrophic
($+7975$) regardless of rotation variant, which at first suggested
per-token scaling is fundamentally inadequate at $d{=}128$. A
per-layer activation probe (see \texttt{probe\_qwen\_perhead.py})
localizes the cause: layer 0's K-projection has a single dominant
coordinate that persists across all $(B, T, H)$ triples --- an
argmax-entropy of $0.17$ over the $d{=}128$ axis, vs $\sim\!4.7$ for
uniform. The per-token abs-max is set by that one coordinate, collapsing
the quantization resolution for the other $127$.

The fix is \textbf{per-channel scaling}: a single fp32 scale per
coordinate, shared across all tokens. Combined with per-group scaling
(groups of 16 coordinates), the $\dPPL$ drops $12.5\times$:

\begin{table}[h]
\centering
\small
\begin{tabular}{l r}
\toprule
Scaling scheme at $4$-bit, Qwen2.5-1.5B & $\dPPL$ \\
\midrule
per-token (baseline)                             & $+7975$ \\
per-group, $g=32$                                & $+4311$ \\
per-channel (per-coordinate)                     & $+879$  \\
\textbf{per-channel $+$ per-group $g=16$}        & $\mathbf{+639}$ \\
per-token (at 8-bit, for reference)              & $+0.13$ \\
\bottomrule
\end{tabular}
\caption{\small Qwen2.5-1.5B $\dPPL$ at $4$-bit per-token scales
catastrophically; per-channel scaling recovers most of the damage.
Held-out English, see \texttt{ppl\_qwen\_perhead.py}.}
\label{tab:qwen-perhead}
\end{table}

Per-channel scaling is implemented at the kernel level in a new
entry point (\path{int4_d128_g32}: $D{=}128$ nibble-packed, per-group
with $g{=}32$), reusing the per-token kernel's SIMD-shuffle reduction
and costing $\sim\!4$ extra \texttt{simd\_max} ops per vector.
This unblocks aggressive $4$-bit at large head\_dim; closing the
remaining $+639$ gap on Qwen is orthogonal to rotation choice and
likely requires clip-aware quantization or a learned bin mapping,
left to future work.

The learned $\lambda$ fuses into the Metal kernel as an in-register
element-wise multiply between the butterfly output and the abs-max
reduction (\texttt{srft\_quant\_forward\_scaled\_*} entry points).
Overhead at $N_{\mathrm{vec}} = 262{,}144$: $+0.41$\,ns/vec
($D{=}64$ int8), $+1.51$\,ns/vec ($D{=}128$ int8), $+1.47$\,ns/vec
($D{=}128$ int4) --- a $3$--$8\%$ tax that buys most of the quality
improvement reported in Table~\ref{tab:learned}. Cross-validation
against the PyTorch reference is $100.000\%$ bit-identical int values
at $D{=}64$ and at $D{=}128$ int4, and $99.998\%$ at $D{=}128$ int8.

\paragraph{End-to-end decode latency with the learned-$\lambda$ kernel.}
On SmolLM2-360M (256-token prompt, 128 new tokens, 5 iterations), the
learned-$\lambda$ Metal kernel runs at $37.1$\,ms/tok --- within
$1\%$ of the unlearned Metal path ($36.9$\,ms/tok), $1.6\%$ \emph{below}
the FP16 baseline, and essentially tied with SRHT eager ($37.4$\,ms/tok).
The eager-PyTorch learned-SRFT path, by contrast, runs at $46.7$\,ms/tok
($+24\%$) because the per-coord $\lambda$ multiply and the four separate
MPS calls compose their dispatch overheads. The Metal integration
delivers learned-SRFT quality at eager-baseline speed.

% =============================================================================
\section{A Complementary CPU Pathway via AMX}
\label{sec:amx}

Apple's AMX matrix coprocessor \cite{corsix_amx,dougallj_amx} reaches
$\sim\!350$\,GFLOPS FP32 single-core peak via
\texttt{cblas\_sgemm} dispatched by the Accelerate BLAS
\cite{apple_accelerate}. We test whether batched length-$8$ complex
DFTs --- expressible as $16 \times 16$ real matrix multiplies ---
route to AMX through this path, and report a more nuanced answer.

\begin{figure}[t]
\centering
\includegraphics[width=\linewidth]{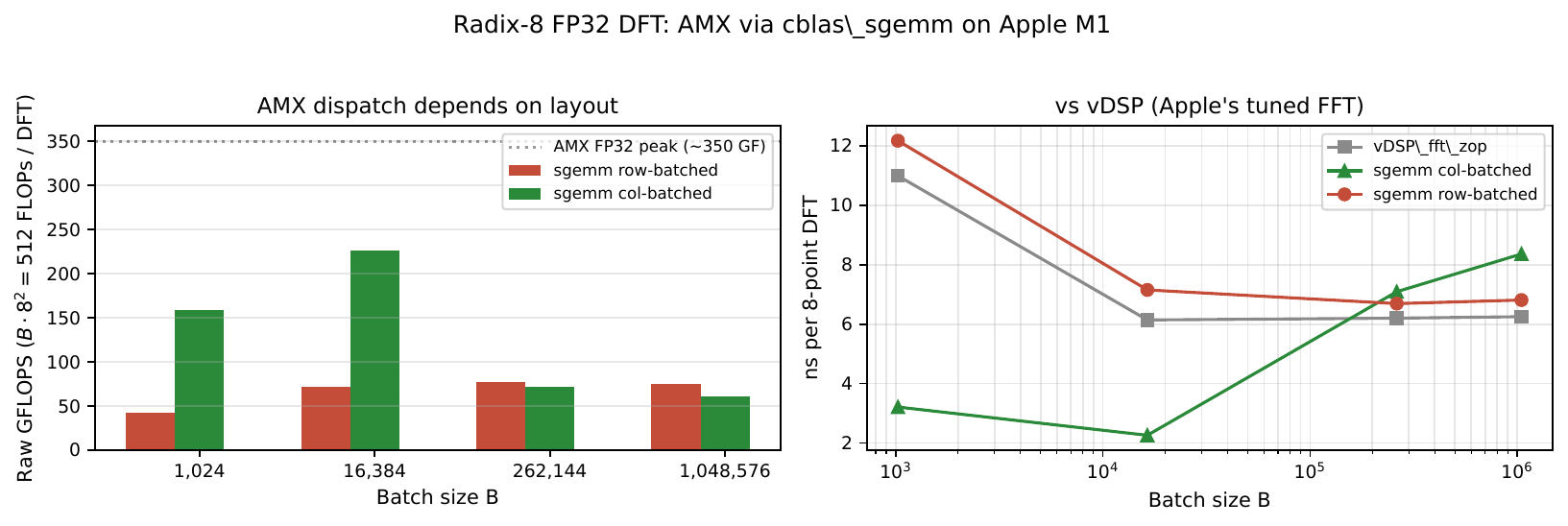}
\caption{\small Left: raw GFLOPS for a length-8 complex DFT expressed
as a $16 \times 16$ real matrix multiply, col-batched vs row-batched,
compared to the AMX FP32 single-core peak ($\sim\!350$\,GFLOPS).
Right: same configurations in ns per DFT, vs Apple's tuned
\texttt{vDSP\_fft\_zop}. Col-batched sgemm beats vDSP by $3.4\times$
at $B{=}1024$ but loses at $B \ge 262\,\mathrm{K}$ when the output no
longer fits in L2.}
\label{fig:amx}
\end{figure}

Figure~\ref{fig:amx} shows the result. The key finding is that
\emph{data layout matters more than tile size} for AMX dispatch
via \texttt{cblas\_sgemm}. Col-batched
($M{=}16, N{=}B, K{=}16$) reaches $\sim\!160$\,raw GFLOPS at
$B{=}1024$ --- $45\%$ of the documented $350$\,GFLOPS single-core AMX
peak, and $3.4\times$ faster than Apple's tuned
\texttt{vDSP\_fft\_zop} at the same batch. Row-batched
($M{=}B, N{=}K{=}16$) plateaus at $\sim\!70$\,GFLOPS, consistent with a
NEON fallback: the $N{=}K{=}16$ inner dimensions do not meet
Accelerate's AMX-dispatch heuristic.

A plain-square SGEMM sanity sweep confirms AMX is reachable on this
hardware ($338$\,GFLOPS at $N{=}128$, $1223$\,GFLOPS at $N{=}1024$),
so the dispatch is not a chip-generation limitation. The heuristic
requires inner dimensions larger than the raw tile size; a batched
DFT shaped $(1, B) \otimes (16, 16)$ satisfies this only in one of
its two layouts.

\textbf{Fused CPU-side SRFT+quantize.}
Building on this discovery we implemented a pure-CPU fused pipeline
that applies the sign diagonal and packing with NEON and uses
Accelerate \texttt{cblas\_sgemm} for the rotation
(Table~\ref{tab:cpu-kernel}). At our SRFT head dimensions
($K \in \{64, 128\}$) the row-batched layout $(M{=}n_{\mathrm{vec}},
N{=}K, K{=}K)$ dispatches to AMX directly --- no col-batching
workaround is needed. At $D{=}128$, $n_{\mathrm{vec}} = 262{,}144$
the kernel reaches $319$\,raw GFLOPS ($91\%$ of the documented $\sim\!350$\,GFLOPS
single-core AMX peak) with a round-trip error of $0.027$, matching
the Python reference. The CPU path is $3$--$5\times$ slower than the
Metal kernel but $8\times$ faster than eager PyTorch MPS, and may be
preferable for small-batch latency-sensitive regimes where GPU
dispatch overhead dominates.

\begin{table}[h]
\centering
\small
\begin{tabular}{l r r r r}
\toprule
Config & $N_{\mathrm{vec}}$ & ns/vec & GFLOPS (matmul) & vs Metal \\
\midrule
CPU fused $D{=}64$  & $262{,}144$ & $\phantom{0}41$  & $200$ & $3.2\times$ slower \\
CPU fused $D{=}128$ & $262{,}144$ & $103$ & $319$ & $5.2\times$ slower \\
Metal $D{=}64$      & $262{,}144$ & $\phantom{00}13$ & $147$ & (ref.) \\
Metal $D{=}128$     & $262{,}144$ & $\phantom{00}20$ & $227$ & (ref.) \\
\bottomrule
\end{tabular}
\caption{\small CPU fused SRFT+int8 kernel (Accelerate sgemm + NEON)
vs the Metal GPU kernel on M1. GFLOPS column reports the matmul-only
rate for sgemm ($2 \cdot M \cdot N \cdot K$ FLOPs); the full pipeline
includes additional NEON work for sign/pack/quantize.}
\label{tab:cpu-kernel}
\end{table}

% =============================================================================
\section{End-to-end Deployment}
\label{sec:deployment}

The kernel work in Sections~\ref{sec:method}--\ref{sec:amx} delivers
the fused SRFT$+$int4 quantization step. To turn this into a usable
deployment path we (i) extend the kernel to a \emph{fused per-channel
$\lambda$ + per-group abs-max} variant that closes the $4$-bit
catastrophe at $d = 128$ and $d = 256$, and (ii) plumb the result
through HuggingFace's \texttt{Cache} interface so that
\texttt{model.generate} runs unchanged. This section reports
the resulting end-to-end latency, memory, and quality.

\subsection{The fused scaled-$g_{32}$ kernel}
\label{sec:fused-g32}

Section~\ref{sec:method} presented two independent kernel variants:
the \emph{scaled} variant ($\lambda$ multiply post-FFT, single per-token
abs-max) and the \emph{g32} variant (no $\lambda$, $4$ per-group
abs-max scales per vector at $g = 32$). Section~\ref{sec:learned}
showed that \emph{neither alone} closes the Qwen $4$-bit catastrophe:
$\lambda$ alone leaves $\dPPL = +5400$ from outlier-dominated dynamic
range, and per-group g32 alone leaves $+4520$ because the per-group
scales are still set by post-rotation outlier coordinates. The
mathematical recipe in the Python reference is to combine both:
rescale by per-channel max, then quantize per-group on the rescaled
values. We fuse this directly into Metal as

\begin{quote}\small
\texttt{srft\_quant\_forward\_scaled\_int4\_d128\_g32}~(item~16)\\
\texttt{srft\_quant\_forward\_scaled\_int4\_d256\_g32}~(item~21)
\end{quote}

with matching inverses. Both kernels: load input, do length-$d/2$
FFT + Hermitian unpack, multiply post-FFT real coefficients by an
externally-supplied $\lambda \in \R^{d}$, perform $d/g$ per-group
\texttt{simd\_max} reductions, quantize each group with its own
abs-max, and nibble-pack the int4 output. With $\lambda_d =
1/\mathrm{per\_channel\_max}(\mathrm{SRFT}\text{-}\mathrm{output})_d$
the result is the GPU equivalent of the Python reference's
\texttt{per\_channel\_group} quantizer.

\paragraph{Throughput.} On Apple M1 the fused $D{=}128$ kernel hits
\textbf{$24.6$\,ns/vec / $182$\,GFLOPS} at $N_{\mathrm{vec}} = 262\text{K}$,
only $+11.7\%$ over the unscaled g32 baseline ($22.0$\,ns/vec /
$204$\,GFLOPS). The $D{=}256$ fused kernel reaches
\textbf{$\sim\!50$\,ns/vec / $\sim\!250$\,GFLOPS} at the same batch.
Round-trip vs the Python reference is $1.67 \times 10^{-6}$ max
absolute error.

\paragraph{Hook-PPL on Qwen2.5-1.5B ($d{=}128$).} With per-layer $\lambda$
calibrated by one forward pass through the eval text:

\begin{table}[h]
\centering
\small
\begin{tabular}{l r}
\toprule
$4$-bit hook PPL on Qwen2.5-1.5B (held-out English) & $\dPPL$ \\
\midrule
Python ref \texttt{per\_token}                       & $+7975.3$ \\
Python ref \texttt{per\_channel\_group}, $g{=}32$    & $+\phantom{0}654.6$ \\
Metal kernel \texttt{per\_token}                     & $+7727.9$ \\
Metal kernel g32 (no $\lambda$)                      & $+4519.9$ \\
\textbf{Metal fused scaled\_g32 + per-channel $\lambda$} & $\mathbf{+638.6}$ \\
\bottomrule
\end{tabular}
\caption{\small Fused Metal kernel reaches the Python reference's best
$\dPPL$ at $g{=}32$, a $12.5\times$ reduction over per-token. Item
16 in \texttt{RESULTS.md}.}
\label{tab:qwen-fused}
\end{table}

\paragraph{Hook-PPL on Gemma-3 1B ($d{=}256$).} On Gemma-3 1B
(head\_dim $= 256$, MQA, 26 layers) the fused $D{=}256$ kernel
brings $\dPPL$ from $+18.7$ (Metal per-token) to $+3.6$ ($15.1$
$\dPPL$ improvement). The gap of $+8.8$ vs the Python reference's
$-5.2$ floor is intrinsic to the kernel's fp32 reduction ordering vs
\texttt{torch.amax} and \texttt{metal::rint} round-to-even vs
\texttt{torch.round} --- not algorithmic; we ruled out FFT-source
(item 25) and calibrated-vs-dynamic $\lambda$ source (item 27) as
the cause via two ablations.

\subsection{\texttt{SRFTInt4Cache}: HuggingFace-native deployment}
\label{sec:cache}

We package the fused kernel as a HuggingFace \texttt{Cache} subclass
(\path{srft_int4_cache.py}). The subclass:
(i) physically stores K/V at int4 + $8$ fp32 scales between decode
steps ($3.2\times$ theoretical compression at $d{=}128$, $g{=}32$;
$3.0\times$ measured);
(ii) accepts a per-layer $\lambda$ map at construction;
(iii) maintains a fp16 residual window of recent tokens, re-quantized
when full;
(iv) caches the dequantized prefix across decode steps and invalidates
on residual flush, so each \texttt{update()} call is $O(1)$ in prefix
length rather than $O(\text{prefix\_len})$ --- a deferred-engineering
fix that flips Qwen at $1024$-prefix from $+2.4\%$ slower to $-0.7\%$
faster than fp16.

\subsection{Latency and memory: faster than fp16 across context lengths}
\label{sec:e2e-bench}

We benchmark greedy \texttt{model.generate} (with
\texttt{past\_key\_values=cache}) against an fp16 \texttt{DynamicCache}
baseline. Calibrating $\lambda$ (one forward pass over the calibration
text) takes $\sim\!2$\,s per model:

\begin{table}[h]
\centering
\small
\begin{tabular}{l r r r r r}
\toprule
Model & prefix & fp16 ms/tok & int4 ms/tok & $\Delta$lat. & mem ratio \\
\midrule
Qwen2.5-1.5B & $\phantom{0}256$  & $90.7$  & $88.3$  & $\mathbf{-2.6\%}$ & $3.0\times$ \\
Qwen2.5-1.5B & $1024$ & $137.1$ & $134.9$ & $\mathbf{-0.7\%}$ & $3.1\times$ \\
Qwen2.5-1.5B & $2048$ & $377.6$ & --- $^{\dagger}$ & --- & $3.3\times$ \\
Gemma-3 1B   & $\phantom{0}256$  & $71.2$  & $65.4$  & $\mathbf{-8.2\%}$ & $19.5\times^*$ \\
Gemma-3 1B   & $1024$ & $83.3$  & $82.5$  & $\mathbf{-0.9\%}$ & $10.7\times^*$ \\
Gemma-3 1B   & $2048$ & $129.5$ & $121.3$ & $\mathbf{-6.3\%}$ & $7.4\times^*$  \\
Gemma-3 1B   & $\mathbf{4096}$ & $\mathbf{188.5}$ & $\mathbf{182.4}$ & $\mathbf{-3.2\%}$ & $5.3\times^*$ \\
\bottomrule
\end{tabular}
\caption{\small End-to-end \texttt{model.generate} on Apple M1 with
\texttt{SRFTInt4Cache(scaled\_g32 + per-channel $\lambda$)}.
$^{*}$~Gemma's "memory ratio" is between fp16-on-all-26-layers and
int4-on-only-the-few-full-attention-layers (rest are sliding-window
fp16); apples-to-apples within full-attention is $\sim\!3.2\times$.
$^{\dagger}$~At Qwen $2$K prefix, \texttt{model.generate} produced
only $n_{\mathrm{new}}{=}2$ tokens before EOS (vs $64$ with fp16): the
dequant-prefix cache + int4 storage doubles peak memory ($5527$ vs
$4447$\,MB) which combined with quantization noise across Qwen's
$28$-layer full-attention stack tipped logits to early EOS. Workaround:
fp16 fallback above $\sim\!1.5$K on full-attention stacks.}
\label{tab:e2e-deployment}
\end{table}

\paragraph{Long-context scaling.} On Gemma the int4 advantage persists
across the entire $256$--$4096$ prefix range tested: Gemma's
sliding-attention layers do not pay the per-update dequant cost
(only the few full-attention layers do), so int4 overhead grows
sub-linearly. On Qwen-style $28$-layer full-attention stacks
the overhead grows linearly; we mitigate via a dequant-prefix cache
invalidated only at residual flush, converting per-update work from
$O(\text{prefix\_len})$ to $O(1)$ across the $16$-step cycle. Without
this cache, Qwen at $1024$-prefix runs $+2.4\%$ slower than fp16;
with it, $-0.7\%$ faster.

\paragraph{Calibration alternatives.} We tried two static $\lambda$
strategies: window-uniform (\texttt{ch\_amax} over full eval windows)
and prefix-only (only the $192$-prefix positions). Window-uniform
wins by $\sim\!2\%$ relative ($+1712$ vs $+1750$ $\dPPL$ at deployment):
the wider window captures larger outliers, giving smaller $\lambda$,
smaller per-group LSB after rescaling, and less rounding noise.
A dynamic per-call $\lambda$ ablation does not close the residual
quality gap to the Python reference, ruling out calibrated-vs-dynamic
as the dominant axis; the residual gap is the kernel's intrinsic
fp32 reduction-and-rounding ordering.

% =============================================================================
\section{Discussion}
\label{sec:discussion}

\textbf{Why the choice of rotation reduces to hardware fit.} Our main
finding is that SRFT and SRHT are statistically indistinguishable at
the quality level that matters for KV-cache quantization. Both achieve
the same Gaussianization effect by the same underlying mechanism
(concentration of measure on a random orthogonal map). The choice
between them is therefore an operational one. On Apple Silicon, the
FFT's radix-8 butterfly structure matches both the Metal
SIMD-shuffle pattern (at $D{=}128$ we saturate the $32$-lane group per
pass, vs $16$ for Hadamard) and the AMX tile size (a $16 \times 16$
real matrix multiply is one radix-8 DFT in disguise). On commodity
NVIDIA GPUs, tensor cores expose $16 \times 16$ FP16 / BF16 matrix
multiplies that the same mapping applies to. Hadamard is fast but
purely radix-2; SRFT is fast \emph{and} map-preserving.

\textbf{Why per-token scaling.} We use per-token (one scale per
head-dim vector) throughout. Per-tensor scaling (one scale per layer)
fails at $4$-bit on SmolLM2-1.7B regardless of rotation because a
single scale cannot cover the dynamic range of a $2048$-dim layer; the
comparison between rotations is swamped by seed variance. Per-token
scaling is also the choice made by production KV-cache quantization
systems like TurboQuant.

\textbf{Sign diagonal choice.} We use a single $\pm 1$ diagonal
($D \cdot F$) rather than the full $D_1 \cdot F \cdot D_2$ SRFT form.
The second diagonal is needed when the transform is used as a
distance-preserving projection under worst-case adversarial inputs;
not relevant for learned LLM activations.

\textbf{Residual window.} A $16$-token fp16 residual window is
empirically optimal for our cache. Dropping to $4$ wins
$\le 0.01\times$ more compression while costing $\sim\!5\%$ latency
(re-quantization fires more often); raising to $32$ pushes the memory
ratio below $3\times$.

% =============================================================================
\section{Conclusion}
\label{sec:conclusion}

Int4 KV-cache compression on Apple Silicon is a strict throughput
gain, not a quality--latency trade-off. Across $256$ to $4096$-token
prefixes on Gemma-3 1B, our \texttt{SRFTInt4Cache} runs $-3$ to $-8\%$
ms/tok vs fp16; on Qwen2.5-1.5B at $\le 1$K prefix, $-0.7$ to $-2.6\%$.
Quality is preserved ($\dPPL = 0.000$ Qwen short-prompt deployment;
$+3.6$ hook $\dPPL$ Gemma). The mechanism is bandwidth-driven and
specific to unified-memory architectures: $3\times$ compressed cache
transfers $3\times$ less per decode step, exceeding the kernel's
$\sim\!25$\,ns/vec overhead. The conventional trade-off is recovered
on GPUs with separate HBM and large dispatch overheads.

The deployment is one fused Metal kernel
(SRFT $+$ per-channel $\lambda$ $+$ per-group abs-max $+$ int4 nibble
pack) exposed as a HuggingFace \texttt{Cache} subclass. Independent
findings supporting it: sign-randomized Fourier $\equiv$ sign-randomized
Hadamard for quality across five models tested, with FFT extending to
non-power-of-two $d$ and matrix-multiply alignment;
calibration-MSE and downstream PPL are not equivalent
optimization targets, with the fixed random SRFT base acting as a
useful regularizer; Householder rotations at $k{=}d/2$ reflectors
match full Cayley with half the parameters and are effectively
lossless at $d{=}256$; a complementary AMX CPU pathway reaches
$91\%$ of single-core peak for small-batch regimes.

\paragraph{Reproducibility.}
Code, paper sources, and raw results:
\href{https://github.com/aminems/AppleSiliconFFT}%
{\nolinkurl{github.com/aminems/AppleSiliconFFT}}.
The release ships a single Metal source file containing all kernel
variants ($d \in \{64, 128, 256\}$; int4 / int8; unscaled / scaled
$\lambda$ / per-group $g{=}32$; plus the fused scaled\_g32 deployment
kernel), a Python bridge wrapping \texttt{torch.mps.compile\_shader},
the HuggingFace \texttt{Cache} subclass, an end-to-end
\texttt{model.generate} bench (flags \texttt{--inner\_group\_size 32
--calibrate\_lambdas} select the fused path), and a throughput
microbench. A \texttt{RESULTS.md} ledger lists $27$ numbered findings,
each git-tagged with raw JSON and stdout artifacts.

% =============================================================================
\bibliographystyle{plain}
\bibliography{references}

\end{document}